\documentclass[usenatbib]{mn2e}

\usepackage{graphicx}
\usepackage{lscape}

\def\a{$^{\mbox{\small a}}$}
\def\b{$^{\mbox{\small b}}$}

\title[V0332+53 in the outburst of 2004--2005: luminosity dependence of the
cyclotron line and pulse profile]{V0332+53 in the outburst of 2004--2005: luminosity dependence of the cyclotron line and pulse profile}
\author[S. S. Tsygankov, A. A. Lutovinov, E. M. Churazov and R. A. Sunyaev]{S. S. Tsygankov$^{1,2}$\thanks{E-mail:st@hea.iki.rssi.ru}, A. A. Lutovinov$^{1,2}$, E. M. Churazov$^{1,2}$ and R. A. Sunyaev$^{1,2}$\\
$^{1}$Space Research Institute, Profsoyuznaya str. 84/32, Moscow 117997, Russia\\
$^{2}$MPI for Astrophysik, Karl-Schwarzschild str. 1, Garching, 85741, Germany}
\begin{document}

\date{Accepted .... Received ...}

\pagerange{\pageref{firstpage}--\pageref{lastpage}} \pubyear{2006}

\maketitle

\label{firstpage}

\begin{abstract}
We present results of observations of the transient X-ray pulsar
V0332+53 performed during a very powerful outburst in Dec, 2004 --
Feb, 2005 with the \textit{INTEGRAL} and \textit{RXTE} observatories
in a wide ($3-100$ keV) energy band. A cyclotron resonance scattering
line at an energy of $\sim26$ keV has been detected in the source
spectrum together with its two higher harmonics at $\sim50$ and
$\sim73$ keV, respectively. We show that the energy of the line is not
constant but linearly changes with the source luminosity. Strong pulse
profile variations, especially near the cyclotron line, are revealed
for different levels of the source intensity. We discuss the obtained
results in terms of the theoretical models of X-ray pulsars.

\end{abstract}


\section{Introduction}

The transient X-ray pulsar V0332+53 was discovered by the Vela 5B observatory
in 1973 \citep{tp73} during an outburst when its intensity reached
$\sim1.4$ Crab in the $3-12$ keV energy band. The outburst lasted
about three months before the source became undetectable again.

During observations in Nov 1983 -- Jan 1984 with the EXOSAT
observatory the pulsar's parameters and orbital ones were determined: the
pulse period $\sim4.375$ s, the orbital period -- $34.25$ days, the
eccentricity -- 0.31, and the projected semimajor axis of the neutron
star -- $a_{x}$sin$i\simeq48$ lt-s \citep{st85}. These authors also
mentioned that as the source intensity decreased the pulse profile
changed from double to single peaked, which was accompanied by
significant hardening of the source spectrum. Later when the source
was observed by the GINGA observatory the cyclotron resonance
scattering feature with an energy of $E_{cyc}=28.5\pm0.5$ keV was
detected in its spectrum. This energy corresponds to a magnetic field
on the neutron star surface of $\sim3\times10^{12}$ G
\citep{mak90}. Later \citet{mih98} reported measurements of two 
different values of the resonance energy for different levels of the
source intensity.

The next powerful outburst of the source began at the end of 2004
\citep{sw04}. This outburst had been predicted based on the increasing
of the optical brightness of the normal companion which reached its
maximum on 31 Jan 2004 \citep{gb04}. A preliminary analysis of
\textit{RXTE} observations performed on 24-26 Dec 2004 showed that
beside the absorption feature at an energy of $26.34\pm0.03$ keV,
there are two additional similar features in the source spectrum at
energies of $49.1\pm0.2$ and $74\pm2$ keV. These features were
interpreted as the second and third harmonics of the main cyclotron
frequency \citep{cob05}. The similar results were inferred from
analysis of the first $\sim100$ ks data of the \textit{INTEGRAL}
observatory \citep{kr05}.

\citet{neg99} discussed the results of the optical observations of BQ Cam
-- the normal companion of the X-ray pulsar V0332+53. They determined
the star's spectral class as O8-9Ve and estimated the distance to the
system at $\sim7$ kpc.

In this paper we present results of an analysis of observations of
V0332+53 performed with the \textit{INTEGRAL} and \textit{RXTE}
observatories in Jan--Feb 2005. Our main aim is to study variations of
the source spectrum and pulse profile depending on its intensity.
Results of the spectral and timing analysis based on RXTE data
and independent analysis of the INTEGRAL data can be found in
\citet{pots05} and \citet{mow06}, respectively.  The observations,
instruments and data analysis are described in section 2. In section 3
we demonstrate for the first time that the energy of the cyclotron
resonance scattering feature is not constant but linearly increases
with decreasing luminosity of the source. Section 4 is dedicated to a
detailed study of the pulse profile variations especially near the
cyclotron line. The obtained results are discussed in section 5.

\section{Observations and Data Analysis}

The X-ray pulsar V0332+53 was observed with the \textit{INTEGRAL} and
\textit{RXTE} observatories several times at the end of 2004 --
beginning of 2005. About 130 pointed observations were carried out by
the \textit{INTEGRAL} observatory \citep{win03} during TOO
observations of the source with a total exposure $\sim400$ ks (these
data are public available). Data of the IBIS telescope (the ISGRI
detector, \citet{leb03}) and X-ray monitor JEM-X \citep{lun03} were
used in the analysis. The effective energy bands of these instruments
are $20-200$ and $3-30$ keV, respectively, which makes it possible to
study bright astrophysical objects in a wide energy band.

The image reconstruction was done with the method and software
described in \citet{rev04}. The source spectra in a hard energy band
($>20$ keV) obtained by the IBIS telescope were calculated through
reconstruction of a large number of images of the source in narrow
energy channels and flux extraction of the studied source. The
response matrix for our research have been done based on the standard
INTEGRAL matrix (taken from the OSA package), but the new arf-file was
calculated using the calibration observations of the Crab
Nebulae. Standard calibration tables and procedures (similar to those
implemented in OSA 5.0) were used to correct the energy of ISGRI
events having different rise-time.  The analysis of a large number of
calibration observations for the Crab nebula revealed that this method
has a systematic error in measuring source fluxes about $\sim3$\%. We
included this systematic uncertainty when analyzing spectra with the
XSPEC package.

For the timing analysis on time scales of the order of the source pulse
period and for the analysis of the JEM-X data we used the standard OSA 5.0
software package provided by the \textit{INTEGRAL} Science Data
Center (http://isdc.unige.ch). For the pulse profiles reconstruction
at high energies the first three steps of OSA (COR, GTI, DEAD) were
executed. After this we collected photons with specified energies
from the detector pixels opened for the source using the tool
\textit{evts\_extract}. At the next stage, the arrival times of
collected photons were corrected for the neutron star orbital motion
using the known binary parameters \citep{st85}. Such a relatively
simple approach for collecting photons is justifiable in our case as
the source V0332+53 was a single object in the telescope field of
view. Thus the detector collects only photons from the source and
background photons. The background can be calculated from those pixels
which are fully closed for the source by the opaque mask elements. To
reconstruct the pulse profile at low energies (JEM-X data) photons
from the whole detector were collected (the tool \textit{evts\_pick}).
The subsequent timing analysis was done with the FTOOLS package.

\begin{figure}
\includegraphics[width=\columnwidth, bb=15 200 570 700, clip]{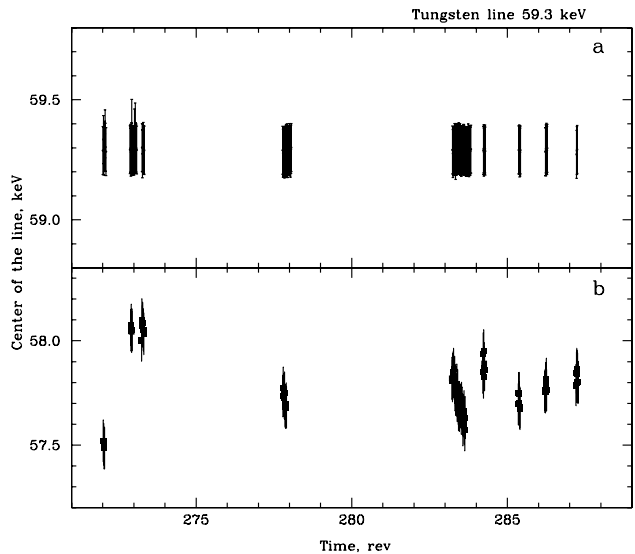}
\caption{Dependence of the energy of the tungsten line $59.3$ keV,
measuring by the ISGRI detector, from the observation date for special
(a) and standard OSA-5.0 (b) softwares.}\label{tung_line}
\end{figure}

Due to a nicety of the considering effects (in particular a
displacement of the cyclotron line center is about $\sim3$ keV in the
INTEGRAL data) and the using of different types of the software on
different stages of the analysis we carried out an additional
investigation of an accuracy and stability of the energy determination
for both type of the software. The tungsten $K_\alpha$ line $59.32$
keV was choose as a calibration line due to its brightness (strictly
speaking, in this spectral region there is a blend of tungten lines at
energies $57.98$, $59.32$ etc. keV, but line at $59.32$ keV is the
most strong). The detector spectrum was built for both softwares (for
OSA 5.0 it was done after the DEAD level), then the energy of the line
center was determined and compared with ones from other observations
and its theoretical value. It was shown that the line center position
is stable and in a good agreement with the theoretical one in the
spectral analysis (Fig.\ref{tung_line}a). The conservative estimation
of the systematic uncertainty of the energy determination gives
$\sim0.1$ keV.

For the standard software OSA 5.0 the line center position is
changed from one observation to another and displace from the
theoretical value (Fig.\ref{tung_line}b). Therefore in the following
analysis we took it into account and made a correction for this
effect.

Apart from \textit{INTEGRAL} data we also used data of the
\textit{RXTE} observatory \citep{br93}: simultaneous data of the
all-sky monitor ASM (http://xte.mit.edu/ASM\_lc.html), and the spectrometers
PCA and HEXTE (Obs.ID 90427), operating in the energy bands $1.3-12.2$,
$3-20$ and $15-250$ keV, respectively. For the \textit{RXTE} data
reduction we used standard programs of FTOOLS/LHEASOFT 5.3 package. As
for the ISGRI detector, we investigated the accuracy of the energy
determination using the HEXTE spectrometer. In this
analysis we used the $^{241}$Am calibration line at $59.6$ keV. To
estimate the systematic error of energy measuring we determined a 
typical scatter of measurements of the line center, $\sim0.1$
keV. This value was added as a systematic error to the uncertainties
obtained from the spectral analysis. 

\begin{table*}
\noindent
\centering
\caption{Observations of the X-ray pulsar V0332+53 with the
\textit{INTEGRAL} and \textit{RXTE} observatories in 2004--2005}\label{tab1} 
\centering
\vspace{1mm}
\small{
\begin{tabular}{l|c|c|c}
\hline\hline
 Date, MJD   & Exposure, & Flux\a,                                 &  Luminosity\b, \\
 (revolution)& ks        & $\times10^{-9}$ erg s$^{-1}$ cm$^{-2}$  &
 $10^{37}$ erg s$^{-1}$ \\
\hline
\multicolumn{4}{c}{}\\[-4mm]
\multicolumn{4}{c}{ \textit{INTEGRAL} data}\\[2mm]
\hline
&&&\\ [-4mm]
53376.5 (272 rev)& 28.7 & $58.3\pm0.9$  &  $34.1\pm0.5$ \\
53379.2 (273 rev)& 56.7 & $45.6\pm0.4$  &  $26.7\pm0.2$ \\
53380.3 (274 rev)& 21.1 & $40.7\pm2.4$  &  $23.8\pm1.4$ \\
53394.0 (278 rev)& 72.3 & $24.9\pm1.0$  &  $14.5\pm0.6$ \\
53410.9 (284 rev)& 149.1& $12.5\pm0.2$  &   $7.3\pm0.1$ \\
53413.1 (285 rev)& 15.5 & $11.4\pm0.3$  &   $6.7\pm0.2$ \\
53416.5 (286 rev)& 15.4 & $ 8.4\pm0.3$  &   $4.9\pm0.2$ \\
53419.1 (287 rev)& 17.0 & $ 7.0\pm0.3$  &   $3.4\pm0.2$ \\
53422.1 (288 rev)& 20.1 & $ 3.5\pm0.2$  &   $2.0\pm0.1$ \\
\hline
\multicolumn{4}{c}{}\\[-4mm]
\multicolumn{4}{c}{\textit{RXTE} data}\\[2mm]
\hline
&&&\\ [-4mm]

53367.2 (90427-01-01-00G)& 1.5 & $82.4\pm2.0$& $48.4\pm1.2$\\
53368.3 (90427-01-01-01) & 2.2 & $81.6\pm2.4$& $48.0\pm1.4$\\
53368.9 (90427-01-01-02) & 1.8 & $79.2\pm0.5$& $46.7\pm0.3$\\
53369.6 (90427-01-01-03) & 1.9 & $72.9\pm0.4$& $42.9\pm0.2$\\
53374.0 (90427-01-02-00) & 0.9& $68.5\pm2.1$& $40.3\pm1.2$\\
53375.0 (90427-01-02-01) & 0.8 & $60.9\pm2.1$& $35.8\pm1.2$\\
53376.3 (90427-01-02-02) & 0.7 & $66.0\pm2.6$& $38.8\pm1.5$\\
53376.7 (90427-01-02-03) & 2.7 & $61.9\pm1.3$& $36.4\pm0.8$\\
53385.1 (90427-01-03-01) & 9.9 & $43.7\pm1.5$& $25.7\pm0.9$\\
53385.5 (90427-01-03-02) & 14.5 & $41.5\pm0.2$& $24.4\pm0.1$\\
53387.0 (90427-01-03-05) & 12.8 & $38.0\pm0.1$& $22.3\pm0.1$\\
53387.4 (90427-01-03-06) & 12.1& $36.5\pm0.1$& $21.5\pm0.1$\\
53387.9 (90427-01-03-07) & 9.6 & $37.2\pm0.1$& $21.9\pm0.1$\\
53388.4 (90427-01-03-09) & 11.9 & $34.5\pm0.8$& $20.3\pm0.5$\\
53389.0 (90427-01-03-11) & 9.9 & $34.5\pm0.4$& $20.3\pm0.2$\\
53389.3 (90427-01-03-12) & 9.7 & $34.1\pm1.0$& $20.1\pm0.6$\\
53413.1 (90427-01-04-00) & 6.1 & $14.7\pm0.6$& $ 8.6\pm0.3$\\
53413.8 (90427-01-04-04) & 6.7 & $12.6\pm0.4$& $ 7.4\pm0.2$\\
53414.3 (90427-01-04-02) & 11.8 & $11.1\pm0.2$& $ 6.5\pm0.1$\\
53414.5 (90427-01-04-03) & 6.8 & $11.0\pm0.3$& $ 6.5\pm0.2$\\
53414.8 (90427-01-04-05) & 2.6 & $10.9\pm0.2$& $ 6.4\pm0.1$\\
53416.6 (90427-01-04-01) & 5.7 & $8.9\pm0.2 $& $ 5.2\pm0.1$\\

\hline
\end{tabular}
\vspace{3mm}

\begin{tabular}{ll}
\a & in the 3-100 keV energy band \\
\b & in the 3-100 keV energy band, assuming source distance of $7$ kpc \\
\end{tabular}}
\end{table*}

A journal of observations of the X-ray pulsar V0332+53 is presented
in Tabl.\ref{tab1}. The date of observations, corresponding exposure,
observed source flux and luminosity are given. The luminosity was
calculated assuming a source distance of 7 kpc. The obtained
luminosity values may be regarded as close to the bolometric ones on
the assumption that the bulk of the energy is released in X-rays.

\begin{figure}
\includegraphics[width=\columnwidth, bb=15 150 570 715, clip]{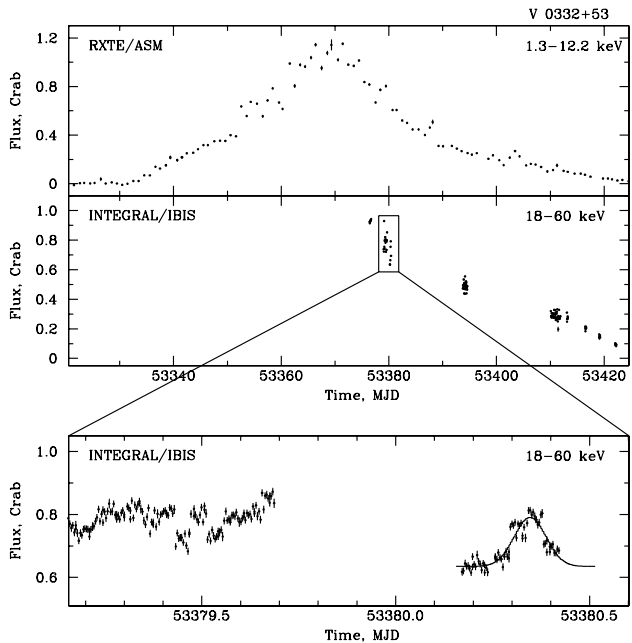}
\caption{Light curves of the X-ray pulsar V0332+53 obtained during 
the outburst: (a) with the ASM/\textit{RXTE} monitor; (b) with the
IBIS/\textit{INTEGRAL} telescope in a hard energy band; (c) with the
IBIS/\textit{INTEGRAL} telescope with a time resolution of 300 s for
the bright state. Solid line represents the best-fit approximation of
the small outburst by a gaussian with maximum at MJD 53380.3 and
total width $\sim4.5$ h.}\label{lcurve}
\end{figure}

\section{Light Curves}

The daily-averaged light curve of the X-ray pulsar V0332+53 obtained
by ASM/\textit{RXTE} is presented in Fig.\ref{lcurve}a. The source
intensity was increasing practically linearly for $\sim30$ days, then
the source stayed for about 10 days near the maximum of its intensity,
and after that the measured flux decreased to the pre-outburst value
in $\sim50$ days. The last segment of the light curve can be described
by an exponential decay with a characteristic time $\tau\sim17$
days. A comparison of the 2004--2005 outburst with previous type
II outbursts \citep{st86} shows that it is a typical outburst 
of this class for V0332+53 in terms of duration and maximum of X-ray
flux.

\textit{INTEGRAL} observations started only several days
after the outburst maximum due to limitations in the satellite
orientation relative to the Sun. In Fig.\ref{lcurve}b the pulsar
light curve obtained by the ISGRI detector in the hard ($18-60$ keV)
energy band is shown. Each point corresponds to a separate
observation with exposure about 2 ks.  The figure demonstrates
that the source flux in hard X-rays decreased from $\sim900$ to
$\sim100$ mCrab during a month and a half. Note that the shape of the
light curve in hard X-rays is slightly different from the one in the
soft $1.3-12.2$ keV energy band.

From the analysis of the V0332+53 light curve in the soft energy band
it was found that the source intensity demonstrates variability with
an amplitude of about 20\% (near the outburst maximum). The
variability amplitude decreased when the source flux decreased and
practically disappeared by the end of the outburst. The variability
with the same amplitude was observed with the ISGRI detector in the
hard energy band too. The 18-60 keV source light curve with time
resolution of 300 s is shown in Fig.\ref{lcurve}c for the bright state
of the source, when the variability is visible most. In the right part
of the Figure a local outburst with maximum at MJD 53380.3 and total
duration about 4.5 hours is seen. The solid line represents the
best-fit approximation by a gaussian.

\section{Spectral Analysis}

The V0332+53 spectrum deserves special attention. This is only the
second X-ray accreting pulsar (after 4U0115+63) whose spectrum exhibit
not only a cyclotron resonance scattering feature but also its two
higher harmonics \citep{cob05,kr05,pots05}. As a whole, the spectrum
of the pulsar can be well described by a power law with an exponential
cutoff at high energies, typical for this class of objects.

The significant number of V0332+53 observations carried out with the
\textit{INTEGRAL} and \textit{RXTE} observatories allowed us to
reconstruct the source spectrum at different phases of the outburst
and to trace the evolution of its parameters. We approximated the
obtained spectra with a powerlaw model with an exponential cutoff
(\textit{cutoffpl} in XSPEC), modified by three absorption
lines in the form

$$exp\left(\frac{-\tau_{cycl}(E/E_{cycl})^2\sigma_{cycl}^2}{(E-E_{cycl})^2+\sigma_{cycl}^2}\right),$$

where $E_{cycl}$, $\sigma_{cycl}$ and $\tau_{cycl}$ are the center,
width and depth of the line, respectively \citep{mih90}. This model
describes the source spectrum similarly well as the usual model of a
powerlaw with a high energy cutoff ($powerlaw\times highecut$ in
XSPEC, White et al. 1983) but has fewer parameters. The best-fit value
of the cutoff energy $E_{cut}$ for the $powerlaw\times highecut$ model
is about $(5-6)$ keV. Given the limited energy range of the used
instruments (down to 3 keV for PCA and 4.5 keV for JEM-X), it is not
possible to put tight constraints on the photon index. An iron
emission line was added to the model when fitting the \textit{RXTE}
data. This feature is not detected by the JEM-X monitor partly due to
its lower sensitivity compared to PCA, and also due to the current
uncertainty in the JEM-X response matrix at these energies (see
comments in
\citet{fil05} for details).

\begin{figure}
\includegraphics[width=\columnwidth, bb=95 355 520 690, clip]{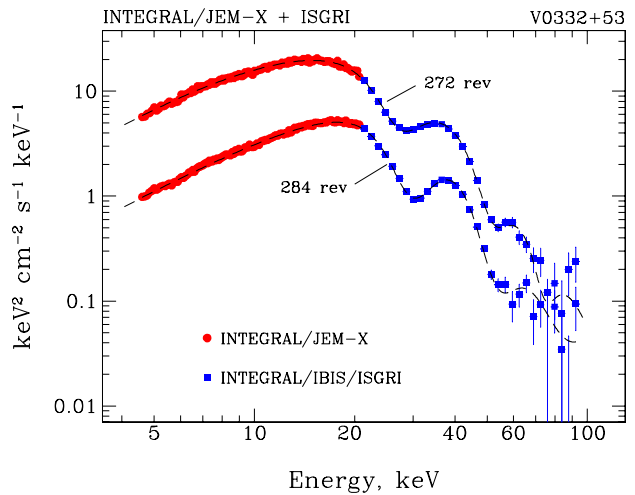}
\caption{Energy spectra of V0332+53 measured with the \textit{INTEGRAL} 
observatory for two states (272 and 284 revolutions).}\label{spectr}
\end{figure}

A typical source spectrum obtained with the \textit{INTEGRAL}
observatory for the bright (272 rev) state is shown in
Fig.\ref{spectr}. The best-fit parameters are summarized below:
\begin{center}
\begin{tabular}{ll}
\hline
Model parameter & Value  \\

\hline

Photon index           & $-0.120\pm0.008$ \\
$E_{cut}$, keV         & $9.21\pm0.04$ \\
$\tau_{cycl,1}$        & $1.91\pm0.02$ \\
$E_{cycl,1}$, keV      & $25.92^{+0.07}_{-0.08}$ \\
$\sigma_{cycl,1}$, keV & $5.44^{+0.08}_{-0.06}$ \\
$\tau_{cycl,2}$        & $2.12\pm0.03$ \\
$E_{cycl,2}$, keV      & $49.44^{+0.07}_{-0.14}$ \\
$\sigma_{cycl,2}$, keV & $9.89^{+0.20}_{-0.23}$ \\
$\tau_{cycl,3}$        & $1.26\pm0.10$ \\
$E_{cycl,3}$, keV      & $72.1^{+0.5}_{-0.6}$ \\
$\sigma_{cycl,3}$, keV & $10.1^{+0.5}_{-0.9}$ \\
$\chi^2$ (d.o.f)       & $1.25 (136)$\\

\hline
\end{tabular}
\end{center}

For comparison, the spectrum obtained in the low state (284 rev) is
shown in the same Figure. 

\begin{figure}
\includegraphics[width=\columnwidth, bb=15 270 520 690,clip]{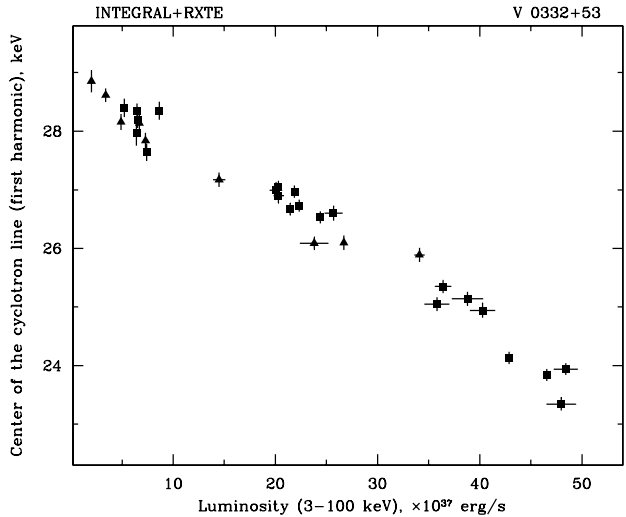}
\caption{The cyclotron line energy dependence on the source luminosity 
(3-100 keV). 
Triangles are \textit{INTEGRAL} results, squares are \textit{RXTE}
ones. }\label{line26}
\end{figure}

The cyclotron resonance scattering feature and its second harmonics
are clearly visible in both spectra. Despite the rapid decrease of the
source intensity and its weakness at high energies ($>65$ keV),
inclusion into the model of the third harmonics with an energy of
$\sim75-80$ keV leads to a significant improvement of the fit ($\Delta
\chi^2=18$ for 3 d.o.f.). This harmonics is also detected in several
next observations, but its parameters (width and depth) are reasonably
bounded only in the bright state (till 284 rev). Fixing them at the
values obtained for the bright state, does not allow one to improve
the quality of the fit for the low-state spectra. Moreover, the
determined line widths for these observations are too large and
strongly affect the determination of the parameters of the second
harmonics, making this task model dependent. It is necessary to note
that the fundamental line energy independs on the including to
the model the third harmonics. It is nessecary to note that
\citet{pots05} used a Gaussian profile for the describing of the
cyclotron line, therefore obtained by them values of the cyclotron
energy are slightly differ from our ones. The same situation was
discussed by \citet{nak06} for 4U0115+634.

Our analysis showed that the model described approximates the
source spectrum well during the whole outburst both for
\textit{INTEGRAL} and \textit{RXTE} data. 
The behaviour of the cyclotron line is of greatest interest because it is
confidently detected during the entire outburst and its parameters are
model independent. The line energy dependence on the source
luminosity obtained from \textit{INTEGRAL} and \textit{RXTE} data is
shown in Fig.\ref{line26} by dark triangles and squares,
respectively. The uncertainties of the HEXTE results are slightly higher
than those for ISGRI as the HEXTE exposures are shorter (see Table
\ref{tab1}). The measurements of both observatories are in good mutual
agreement and fall on a straight line, i.e. the cyclotron line energy
increases linearly with decreasing source luminosity. The formal
fitting of this dependence with a linear relation gives 
$E_{cycl,1}\simeq-0.10L_{37}+28.97$ keV, where $L_{37}$ -- luminosity
in units of $10^{37}$ erg s$^{-1}$. Believing that for low
luminosities the emission come practically from the neutron star
surface (see section 6.1) we can estimate the magnetic field on the
surface 

$$B_{NS}=\frac{1}{\sqrt{1-\frac{2GM_{NS}}{R_{NS}c^2}}}
\frac{28.97}{11.6}\simeq3.0\times10^{12} {\rm G}$$

where $R_{NS}$ and $M_{NS}$ -- are the neutron star radius and mass,
respectively.

\begin{figure}
\includegraphics[width=\columnwidth, bb=15 270 520 690, clip]{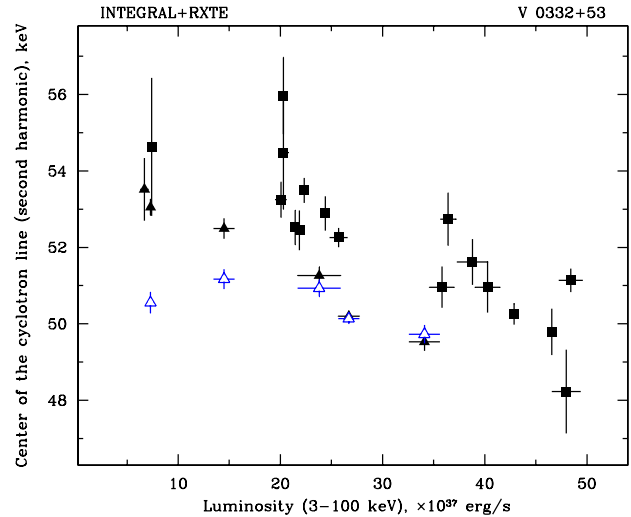}
\caption{The same as in Fig.\ref{line26} but for the second harmonics. 
Open triangles represents \textit{INTEGRAL} measurements for the broad
energy band (till $110$ keV) with inclusion of the third harmonics
(see text).}\label{line51}
\end{figure}

As was mentioned above, the energy of the second harmonics is not
reasonably determined for all observations and depends on the
inclusion of the third harmonics in the model. To avoid possible
contamination by this component in determining of the parameters of
the second harmonics, we restricted the considered energy band by $65$
keV and fitted the spectra by the same model with two absorption
lines. The variation of the second harmonics with source luminosity is
shown in Fig.\ref{line51} by dark triangles and squares similarly to
Fig.\ref{line26}. It can be seen that although the scatter is slightly
larger than for the first harmonics, the overall tendency of the line
energy increasing with decreasing luminosity holds in this case
too. Formal fitting of this dependence with a linear relation gives
$E_{cycl,2}\propto -0.08L_{37}$. The energy of the second harmonics
was also determined by analyzing the \textit{INTEGRAL} data in the
broad energy band (up to 110 keV) when the third harmonics was added
to the model (open triangles in Fig.\ref{line51}). The energies
obtained in both analyses differ from each other, especially for the
observations with low luminosities. Moreover, the ISGRI measurements
lie below the near simultaneous HEXTE ones. This fact is most likely
connected with the larger exposures of the ISGRI detector -- its data
at high energies provide better statistics and the third harmonics
affects the determination of the parameters of the second harmonics
even in the case of the energy band truncated at 65 keV.
If we exclude from
the consideration the \textit{INTEGRAL} results, then formal
fitting of the HEXTE measurements results in $E_{cycl,2}\propto
-0.1L_{37}$, which is the same as was obtained earlier for the main harmonics.

\section{Pulse Profile}

\begin{figure}
\includegraphics[width=\columnwidth, bb=50 150 520 740, clip]{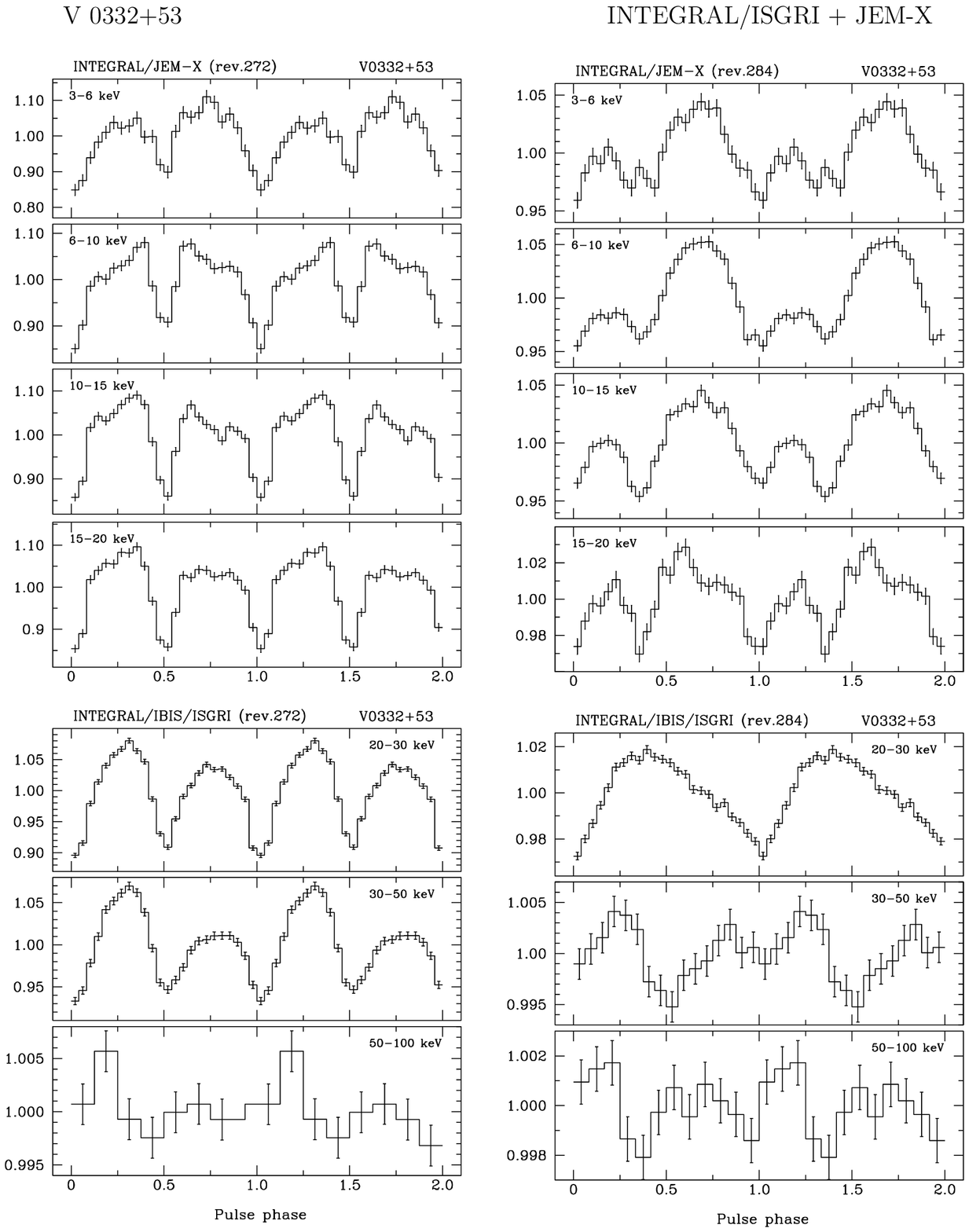}
\caption{V0332+53 pulse profiles obtained with the JEM-X monitor and 
IBIS telescope of the \textit{INTEGRAL} observatory in different
energy bands for different source intensities. The pulse phase zero for
each observation are chosen independently.}\label{pp_tot}
\end{figure}

Because of the high intensity of the pulsar emission we succeeded in
studying its pulse profile dependence on the energy band and
luminosity. As the background doesn't affect on the pulse profile
shape, which is studied in the paper, it was not subtracted in the
following analysis. The most characteristic source pulse profiles
obtained with the \textit{INTEGRAL} observatory in different energy
bands are shown in Fig.\ref{pp_tot} for two observations (272 and 284
revolutions). The observed peculiarities in the pulse profile behavior
can be divided in two main groups: an asymmetrical evolution of the
double-peaked profile in wide energy bands and its drastic changes
near the main harmonics of the cyclotron line. Below we describe these
effects in detail.

\begin{figure*}
\includegraphics[width=0.73\textwidth, bb=50 50 575 790, angle=90,clip]{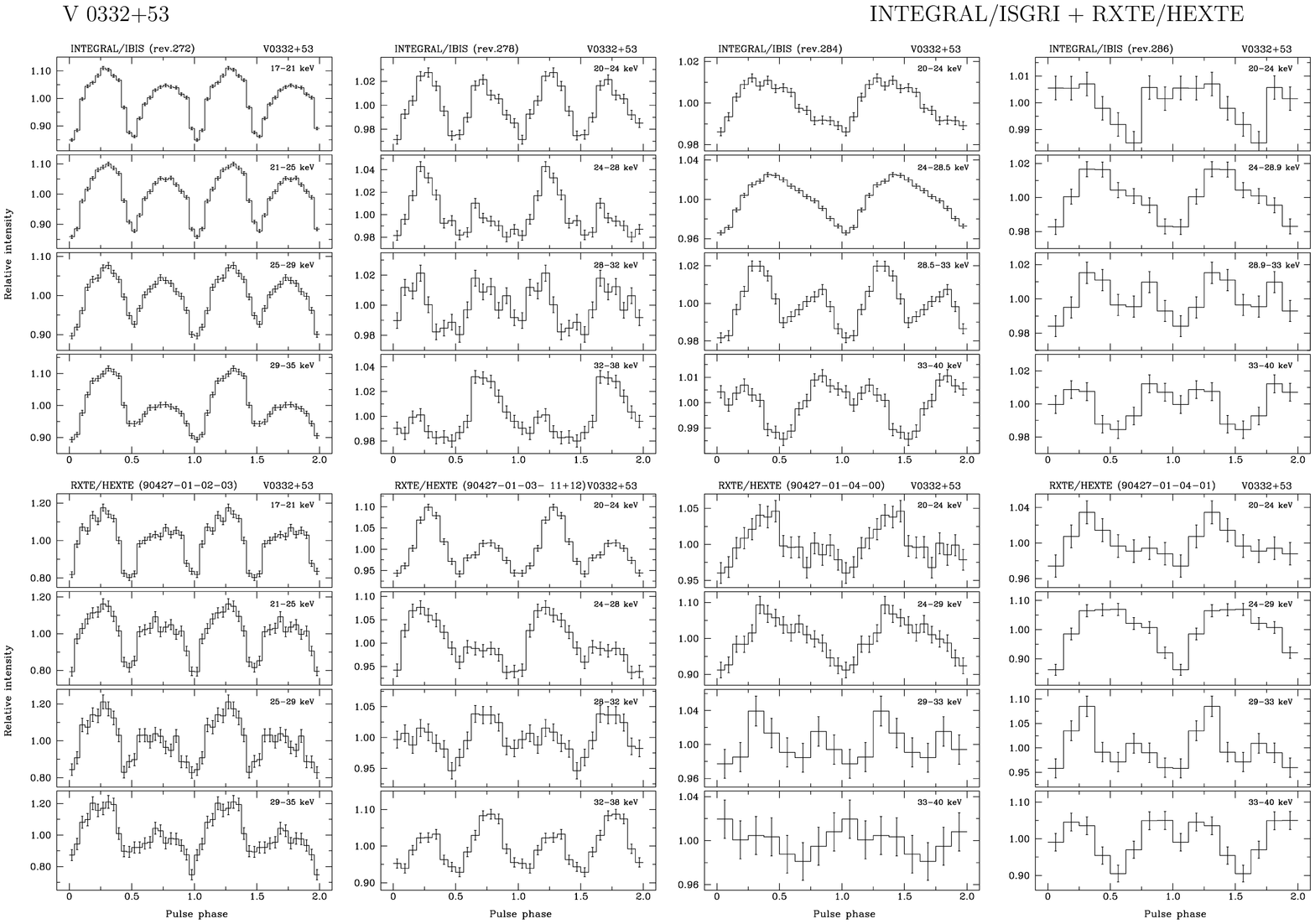}
\caption{Evolution of the source pulse profile near the cyclotron 
line with the source intensity. IBIS/\textit{INTEGRAL} data (upper
panels), HEXTE/\textit{RXTE} data (bottom panels). The pulse phase zero for
each observation are chosen independently.}\label{pp_cycl}
\end{figure*}

In the brightest state (272 rev) the pulse profile has a
sinusoidally-like double-peaked shape with little prevalence of the
second peak at low energies ($3-6$ keV). As the photons energy grows
the relative contribution of the first peak increases and 
exceeds the second peak intensity in the $10-15$ keV energy band.
The intensity of the first peak continues to grow when increasing
the energy and becomes significantly larger than the second one in the
$30-50$ keV channel. No significant movements of the peaks depending on the
energy band occur.

As the pulsar intensity diminishes, significant changes of the pulse
profile occur in soft energy bands (JEM-X data): in the $3-6$ keV band
it becomes nearly single-peaked in the 274 revolutions; the relative
intensity of the first peak is very small in $3-6$ and $6-10$ keV
bands in 278 and 284 revolutions. In high-energy channels (ISGRI data)
the pulse profile evolves similarly as was described above for the 272
revolution. In the following observations (beginning from the 284
revolution when the source luminosity dropped to
$\sim7.3\times10^{37}$ erg s$^{-1}$) the main described tendencies
persist at soft energies, but at energies of order and above the
cyclotron one a drastic change in the pulse profile takes place: the
intra-peak gap disappears and the double-peaked profile transforms
into an asymmetrical single-peaked one. When the energy increases, the
phase of the main minimum is displaced on $\sim0.5$
(Fig.\ref{pp_tot}). The described features remain approximately the
same in wide energy channels for the subsequent observations, i.e. at
lower luminosities.

A detailed study of pulse profiles was carried out in narrow energy
channels in the most interesting region -- around the main harmonics
of the cyclotron line (Fig.\ref{pp_cycl}). Such a study was performed
for four \textit{INTEGRAL} observations (revolutions 272, 278, 284 and
286) and for four \textit{RXTE} observations which are close to
\textit{INTEGRAL} ones (90427-01-02-03, 90427-01-03- 11+12,
90427-01-04-00 and 90427-01-04, respectively). The last was done to
obtain an independent confirmation of the \textit{INTEGRAL}
results. In Fig.\ref{pp_cycl} pulse profiles for these INTEGRAL
observations are drawn. The energy channels were chosen to divide the
cyclotron line in half, their width was approximately the same and
about a half of the cyclotron line width (i.e. two central channels
correspond to the lower and upper wings of the line). Here we took into
account that the energy of the line center changes with the source
luminosity.

For the brightest state (272 rev), the passage through the line
center does not affect the pulse profile and all the tendencies described above
for wide channels remain. In the subsequent observations
(273-278 revolutions) some changes occur in the relative intensity of
the peaks and their positions, but the pulse profile remains
double-peaked. With a further decrease of the luminosity to 
$\sim7.3\times10^{37}$ erg s$^{-1}$ (284 revolution) the profile
became asymmetrical single-peaked at energies below the cyclotron line
with a drastic transition to the double-peaked shape above the line
energy. Unlike in the 278 revolution the displacement of the profile
by half a period, expressed in the displacement of the main minimum, occurs
not in the upper wing of the line but in the next energy channel
(Fig.\ref{pp_cycl}). A similar picture remains when the pulsar
luminosity is decreased to $\sim4.9\times10^{37}$ erg s$^{-1}$ (286
rev). In the next revolution (the luminosity is $\sim3.4\times10^{37}$
erg s$^{-1}$) the pulse profile at energies below the line returned to the
double-peaked shape, although the single-peaked one holds in the lower
wing. The displacement of the profile by half a period is retained in
the last channel too. It is necessary to note that due to the source
intensity decreasing the statistics in the last observations is not
sufficient for detailed analysis of the pulse profile structure and we can
only consider their common characteristics.

The results of a similar analysis performed on the \textit{RXTE}/HEXTE
data (bottom panels on Fig.\ref{pp_cycl}) fully confirm the
conclusions about the behavior of the pulse profile drawn above based
on the \textit{INTEGRAL}/ISGRI data.

\section{Discussion}

The transient X-ray pulsar V0332+53 demonstrates powerful outbursts in
which its intensity exceeds 1 Crab. The source is a member of a high
mass X-ray binary system with a companion star (BQ Cam) that belongs
to the class of $Be$ stars. According to current ideas
(e.g. \citet{oka01}) such objects represent quick rotating stars with
a dense, but radially slow, stellar wind. This wind forms the
so-called equatorial disc around the star, the size and the presence
of the disc are not permanent. Most X-ray sources in binary systems
with $Be$ stars are transients demonstrating outbursting
activity. This is presumably connected with the evolution of the
normal star. Matter from the equatorial disc is captured and an
accretion disc around the relativistic object is formed.

The system V0332+53/BQ Cam is an obvious example of the picture
described above. As was shown by \citet{gor01}, a significant
brightening of the normal star in the optical waveband preceded
previous X-ray outbursts. It is believed that this brightening is
associated with the formation and subsequent ejection of the ambient
matter (equatorial disc) \citep{gor01}. The outburst that started in
Dec, 2004 was no exception and had been predicted based on an increase
in the brightness of the optical star at the beginning of 2004
\citep{gb04}. Such a large (several hundred days) delay between the
optical and X-ray light is typical for this source \citet{gor01} and
is possibly asscociated with the necessity to accumulate in the
accretion disc the sufficient amount of matter to begin accretion. The
idea that accretion during outbursts proceeds through the disc is
supported by the high observed luminosity ($\sim5\times10^{38}$ erg
s$^{-1}$), unachievable for stellar wind accretion. Furthermore, the
pulse period changes at a high rate during the outburst (Tsygankov et
al., in preparation), which is typical for binary systems with disc
accretion and observed in other transient X-ray pulsars with $Be$
companions (see e.g. Tsygankov \& Lutovinov 2005 for KS1947+300).

\subsection{Cyclotron Line}

It was found by \citet{bs76} that there is a critical value of the
luminosity ($L^*\sim10^{37}$ erg s$^{-1}$) dividing two accretion
regimes: the regime when the influence of the radiation on the falling
matter is negligible and the regime when this influence is
significant. When $L<L^*$, the matter free-fall zone is extended
almost down to the surface of the neutron star. In the opposite case
($L>L^*$), observed for V0332+53, the radiation-dominated shock rises
high above the neutron star surface. Almost all of the kinetic energy
of the infalling gas is lost in this shock, and is then emitted
laterally by the sides of the accretion column.

\citet{bs76} and \citet{ls88} showed that the height of the shock $H$ depends
on the source luminosity

$$H\simeq\dot m R_{NS}\ln\left(\eta \frac{1+\dot m}{\dot m^{5/4}}\right)$$

where $\dot m$ is the dimensionless accretion rate in units of
$10^{39}$ erg s$^{-1}$, $\eta$ -- a function depending on the
magnetic field on the neutron star surface $B_{NS}$ and the
thickness of the accretion column. As can be seen from the
equation, the height $H$ changes practically linearly with $\dot m$
in a wide range of values, i.e. the shock height grows linearly when
the source luminosity is increased.
\begin{figure}
\includegraphics[width=\columnwidth, bb=15 270 520 690, clip]{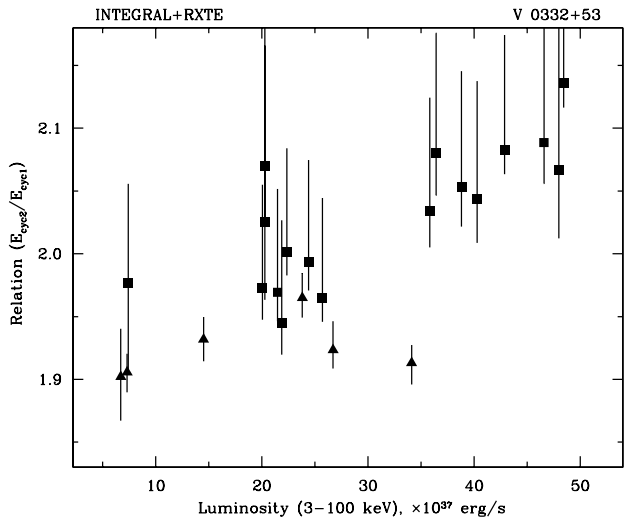}
\caption{Dependence of the ratio of energies of second and main 
harmonics of the cyclotron line on the source luminosity (3-100 keV). Triangles
and squares represents \textit{INTEGRAL} and \textit{RXTE} results,
respectively.}\label{relation}
\end{figure}

It was shown above that the energy of the cyclotron line detected in
the V0332+53 spectrum grows approximately linearly with decreasing
source luminosity. The maximum relative change of the energy and,
consequently, the corresponding magnetic field are about
$\sim25$\%. In an approaching of the dipole field of the neutron star,
it corresponds to a $7.5$\% relative change of the height $h$ where
the feature is formed. At the end of the outburst, the source
luminosity falls to $\sim10^{37}$ erg s$^{-1}$, the shock descends,
the column height decreases and we receive emission coming virtually
from the neutron star surface. Because of the smallness of the
relative change $h/R_{NS}$, we can consider $B(h)\propto B_{NS}-\alpha
h$ to a first approximation, where $\alpha$ is a
coefficient. Comparing with the relation between $E_{cycl,1}$ and $L$
determined earlier we obtain $h\propto L$. Thus the height $h$, where
the cyclotron feature is formed, has the same luminosity behaviour as
the shock height $H$.

\begin{figure*}
\hbox{
\includegraphics[width=\columnwidth, bb=15 50 305 307, clip]{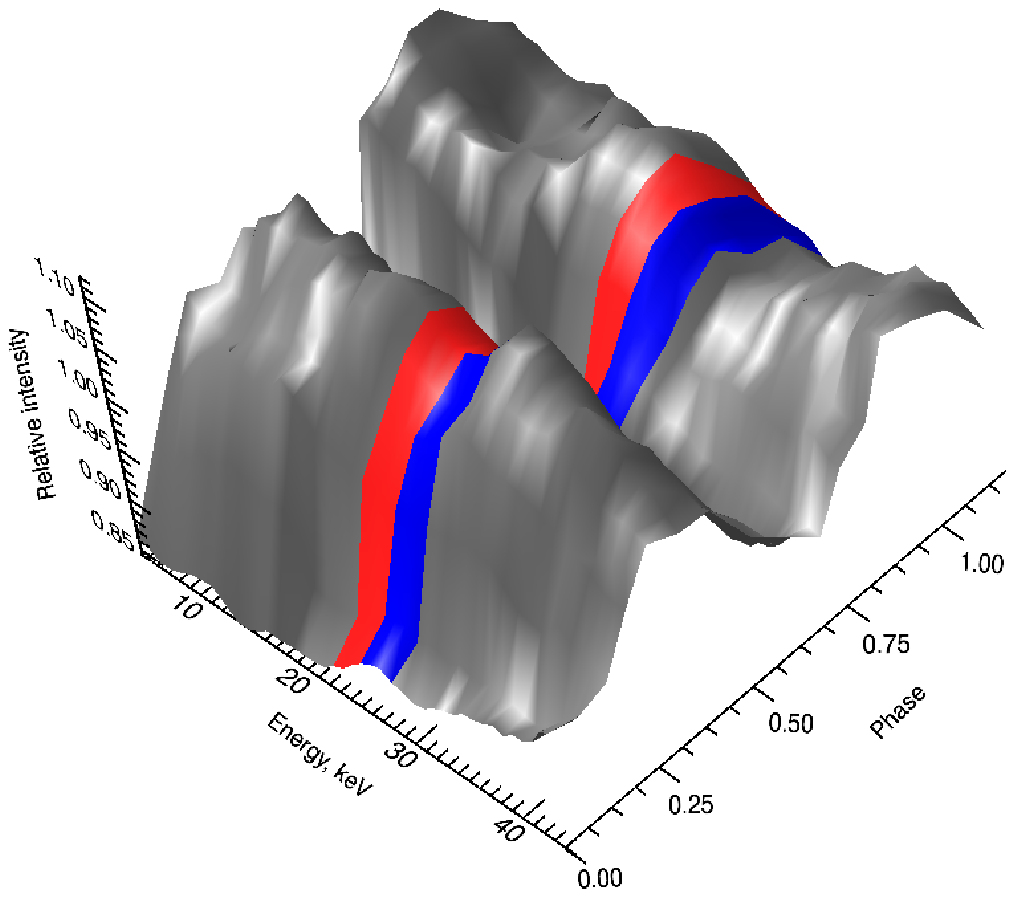}
\includegraphics[width=1.07\columnwidth, bb=25 70 291 280, clip]{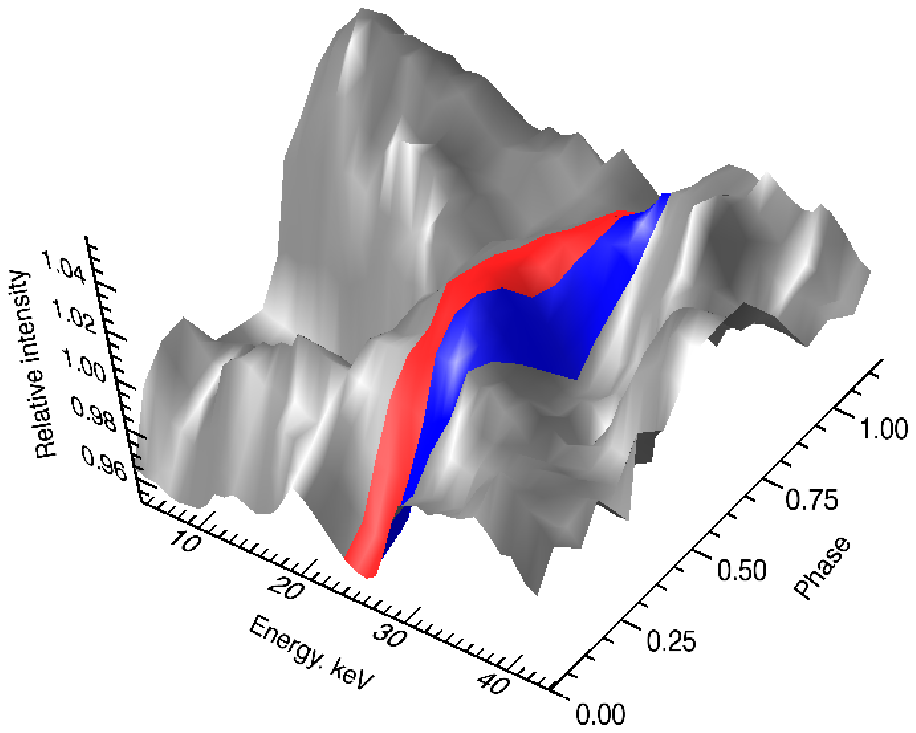}
}
\hbox{
\includegraphics[width=0.97\columnwidth, bb=40 300 530 740, clip]{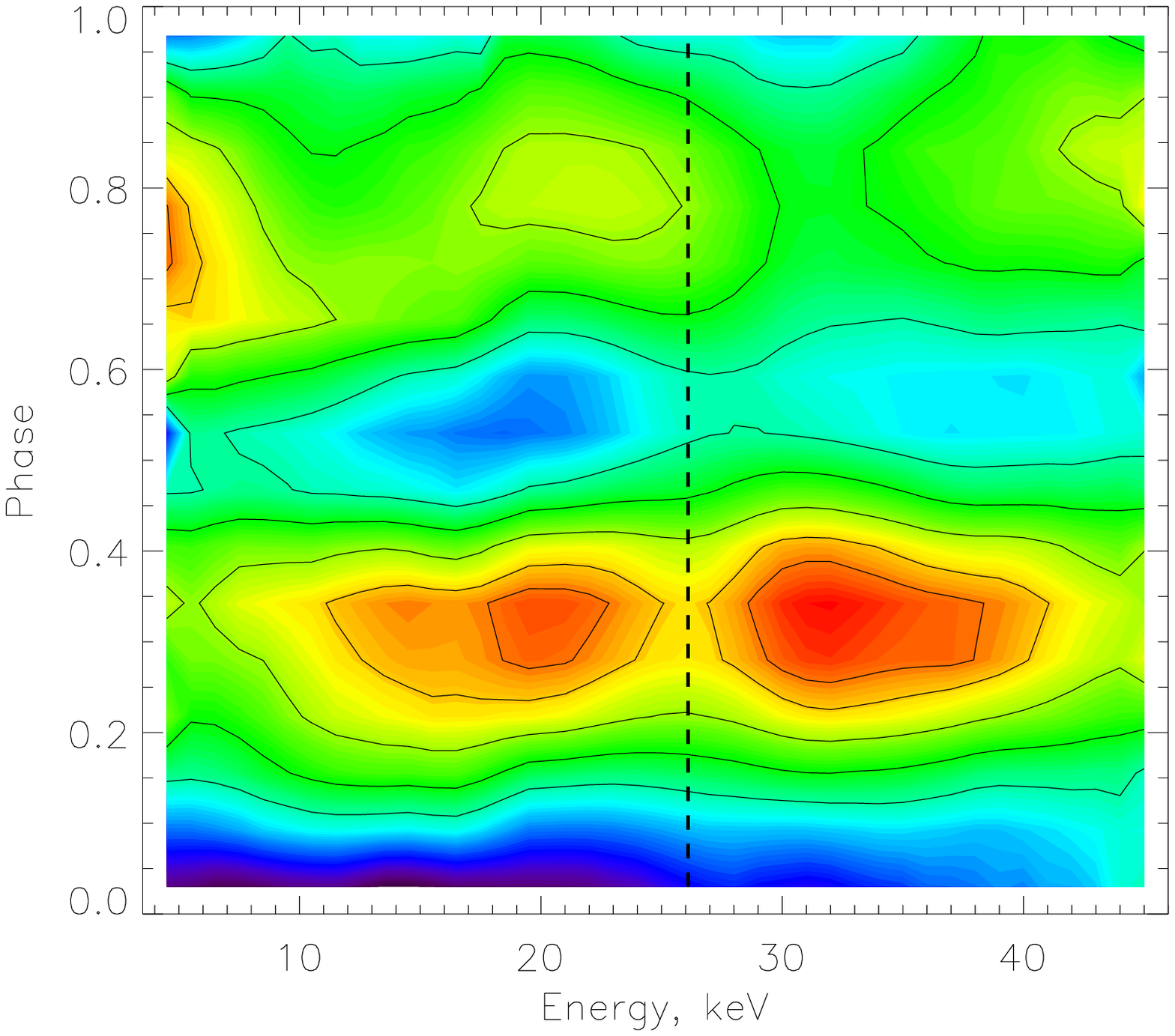}
\hspace{3mm}\includegraphics[width=\columnwidth, bb=20 300 530 740, clip]{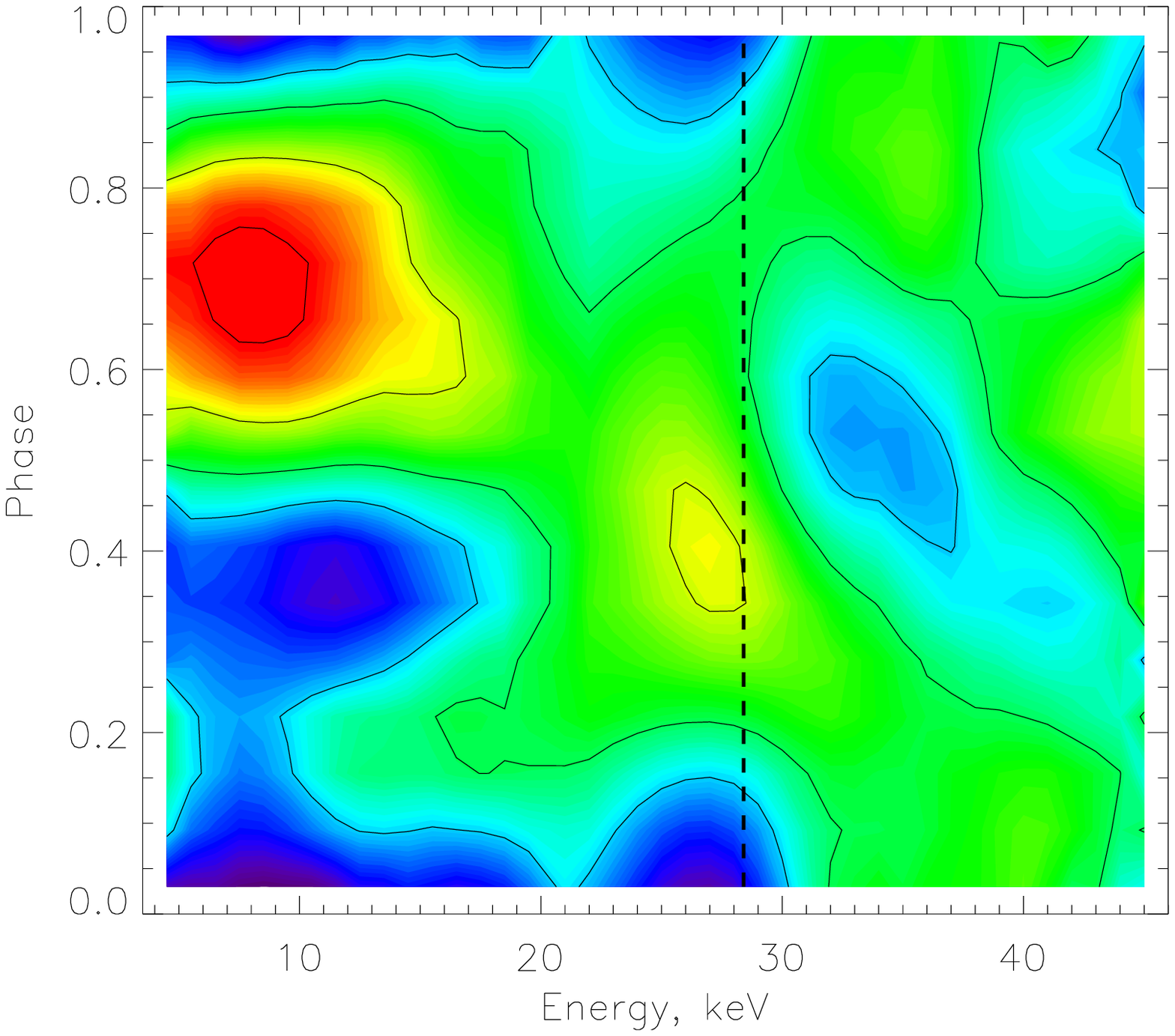}
}
\caption{3D evolution of the source pulse profiles for 272 (a) and 284 
(b) revolutions. Red and blue stripes represents lower and upper wings
of the cyclotron line. 2D-distributions of pulse profile intensities
with the energy and pulse phase are shown in the bottom panel by
different colors. Solid lines represent levels of equal intensity.
Positions of the cyclotron line center are shown by dashed
lines.}\label{3dpp}
\end{figure*}

According to \citet{bs76}, only a small fraction of the energy
accumulated in the accreting matter is emitted at the shock. Its
main part goes into the extended sinking zone below the shock, being
gradually emitted by the side walls of the accreting column.
Moreover, the column cross-section has a near-triangular shape where the
shock proper occupies only a small part in the column central plane.
In other parts the deceleration of the infalling matter is cased by
a friction force \citet{ls88}. Thus, the registered emission is a
superposition of emissions from different heights above the neutron
star surface. Therefore, we can consider the height $h$ as some
averaged or ``effective'' height of the formation of the cyclotron
feature not coinciding with the position of the shock itself.

As seen from Fig.\ref{line51}, the behavior of the energy of the second
harmonics is qualitatively similar to the main one. But due to lower
statistics of the data at high energies the exact determination of its
parameters is model-dependent (see above). Thus at the moment we
cannot make a final conclusion about the rate of change of the second
harmonics energy with the source luminosity.  Nevertheless it is
interesting to note that the ratio of the energies of the second and main
harmonics slightly decreases with decreasing luminosity and is
approximately equal to the harmonic one ($2:1$) near the luminosity
$\sim2\times10^{38}$ erg s$^{-1}$ (Fig.\ref{relation}).

\subsection{Pulse Profile}

As experimental information about pulse profiles and their evolution
has been accumulated, it has become clear that a simple model
explaining pulsations by the presence of two bright spots on the
neutron star surface cannot clarify the variety of observations. Our
results on V0332+53 confirm this statement. 

As was shown above, the matter flowing from the accretion disc forms
near the magnetic poles accretion columns elongated along the magnetic
lines of force. Since the falling matter is opaque, the radiation is
entrained in the column and moves with the matter downward, diffusing
to the edges of the accretion channel and escaping laterally
\citep{ls88}. Thus a fan beam configuration of the X-ray emission is
expected to prevail in the bright state \citep{bs76}, which explains
the observed double-peaked structure of the source pulse profile. But
it is worth noting that although the radiation has time to escape from
the column, it will beam toward the neutron star surface due to
relativistic effects \citep{ls88} and simple considerations about the
beaming can be inapplicable. Most likely the observed pulse profile
variability can be explained by a combination of geometrical and
physical effects. We point out some of them below.

One of the possibilities to describe the observed changes of relative
intensities of peaks in the pulse profile can be the mechanism of 
pulse formation for large source luminosities proposed by
\citet{bs76b}. Its essence is that the magnetic field of the 
neutron star makes the gas to flow off to the magnetic funnels along
the Alfven surface. This flow will cover only a part of the
surface. The layer of matter on the surface will spin with the same
angular velocity as the neutron star and will periodically shield from
the observer different parts of emission regions. For a certain
orientation of the system to the observer it is possible to expect
different relative intensities of peaks in a dependence on the energy
band. Also the reflection from the inner surface of the gas layer
flowing to the lower magnetic field can make a contribution to the
formation of the pulse profile \citep{bs76b}. When the source
intensity is decreased, the scattering optical depth of the masking
layer is also decreased, which can result in a larger amplitude of the
asymmetry in the pulse profile and to an appearance of some new
features in them. The calculations show that in bright states the flows
on the magnetosphere are optically thick. This results in additional
changes of the pulse profile, especially strong in the soft and hard
energy bands \citep{bs76b}.

The most interesting and difficult for explanation changes occur near
the main harmonics of the cyclotron line at luminosities lower
$\sim7.3\times10^{37}$ erg s$^{-1}$. The observed behavior is
difficult to describe in detail within the framework of current
models. In partial, it can be connected with peculiarities of the
radiation beaming near the cyclotron frequency \citep{gs73,pav85}. As
e.g. \citet{mes85} showed, the cyclotron line shape demonstrates a
strong angular dependence. The plasma is more transparent at large
angles than at small ones for energies below and above the line
energy. Therefore photons will escape predominantly in the directions
of large angles, i.e. the radiation beaming in different energy
channels near the cyclotron line will be strongly different. In
addition, a bulk motion towards the neutron star
surface also produces a doppler shift, which should result in an
angular dependence of the cyclotron line energy on the viewing angle,
that in its turn can give a contamination to the observed line energy
and pulse profile dependencies \citet{brai91}.

For the better visual perception and understanding of the changes
described above we built three-dimentional pulse profiles with the
distribution of their relative intensities along the pulse phase and
energy. To obtain a more or less smooth picture we choose the energy
window (the energy band for each profile) of 4 keV and reconstructed a
number of pulse profiles with a step of 1 keV from 6 up to 45
keV. Such an approach allowed us also to sew together JEM-X and ISGRI
results around 20 keV relatively well. Two such 3D pulse profiles for
272 and 284 revolutions are shown in Fig.\ref{3dpp} (upper
panel). Clearly seen all described in section 5 features for both of
them. The red and blue stripes represents regions of lower and upper
wings of the cyclotron lines. In bottom panels of Fig.\ref{3dpp}
two-dimentional distributions of pulse profile intensities are
demonstrated by different colors and levels of equal intensities. It
is interesting to trace changes of the maximum intensities for both
observations: in the first case positions of both peaks are
practically unchanged with the energy; in the second one the position
of the maximum is changed drastically with the energy especially near
the cyclotron line. The single-peaked and double-peaked distribution
of intensities in the lower and upper wings of the cyclotron line are
obviously seen.

The case of small luminosities is of especial interest for following
investigations. In this case the radiating plasma have an appreciable
optical depth only near the cyclotron line and the source pulse
profile reflects physical properties of the plasma flow in the
magnetic field, i.e. is determined by the anisotropy of the emission
and scattering in the plasma.

\section{Summary}

We presented results of the analysis of the \textit{INTEGRAL} and
\textit{RXTE} data obtained during the outburst from the
X-ray pulsar V0332+53. The most important are:\\

-- for the first time we studied in detail the evolution of the
cyclotron energy with the source luminosity and showed that it is
linearly increasing with the source luminosity decreasing in the same
way as the change of the height of the accretion column;

-- the behavior of the second harmonics energy is qualitatively
similar to the main one, but more accurate observations are needed for
exact measurements of its rate and understanding of the behavior
of the ratio of energies of second and first harmonics;

-- the strong pulse profile variations with luminosity, especially
near the cyclotron line, are revealed.

\section*{Acknowledgments}

We thank M.Revnivtsev, R.Krivonos, M.Gilfanov and S.Sazonov for a help
with the data analysis and discussion of the results obtained. We
also thank to the anonymous referee for useful and detailed
comments. This work was supported by the Russian Foundation for Basic
Research (project no.04-02-17276), the Russian Academy of Sciences
(The Origins and evolution of stars and galaxies program) and
grant of President of RF (NSh-1100.2006.2). AL acknowledges financial
support from the Russian Science Support Foundation.  We are grateful
to the European \textit{INTEGRAL} Science Data Center (Versoix,
Switzerland), the Russian \textit{INTEGRAL} Science Data Center
(Moscow, Russia) and the High Energy Astrophysics Science Archive
Research Center Online Service, provided by the NASA/Goddard Space
Flight Center, for the data. The results of this work are partially
based on observations of the \textit{INTEGRAL} observatory, an ESA
project with the participation of Denmark, France, Germany, Italy,
Switzerland, Spain, the Czech Republic, Poland, Russia and the United
States.

\label{lastpage}


\begin{thebibliography}{99}

\bibitem[\protect\citeauthoryear{Basko \& Sunyaev}{1976a}]{bs76}
Basko M.M., Sunyaev R.A., 1976a, MNRAS, 175, 395

\bibitem[\protect\citeauthoryear{Basko \& Sunyaev}{1976b}]{bs76b}
Basko M.M., Sunyaev R.A., 1976b, Sov. Astron., 20, 537

\bibitem[\protect\citeauthoryear{Bradt et al.}{1993}]{br93}
Bradt H.V., Rothschild R.E., Swank J.H., 1993, A\&AS, 97, 355

\bibitem[\protect\citeauthoryear{Brainerd \& Mezaros}{1991}]{brai91} 
Brainerd J., Mezaros P., 1991, ApJ, 369, 179

\bibitem[\protect\citeauthoryear{Coburn et al.}{2005}]{cob05}
Coburn W., Kretschman P., Kreykenbohm I., et al., 2005,
Astron. Telegram, 381, 1


\bibitem[\protect\citeauthoryear{Filippova et al.}{2005}]{fil05}
Filippova E., Tsygankov S., Lutovinov A., Sunyaev R., 2005,
Astron. Lett., 31, 729

\bibitem[\protect\citeauthoryear{Gnedin \& Sunyaev}{1973}]{gs73}
Gnedin Yu., Sunyaev R., 1973, A\&A, 25, 233

\bibitem[\protect\citeauthoryear{Goranskij}{2001}]{gor01}
Goranskij V., 2001, Astron. Lett., 27, 516

\bibitem[\protect\citeauthoryear{Goranskij \& Barsukova}{2004}]{gb04}
Goranskij V., Barsukova E., 2004, Astron. Telegram, 245, 1

\bibitem[\protect\citeauthoryear{Kreykenbohm et al.}{2005}]{kr05}
Kreykenbohm I., Mowlavi N., Produit N., et al., 2005, A\&A, 433, L45

\bibitem[\protect\citeauthoryear{Lebrun et al.}{2003}]{leb03}
Lebrun F., Leray J. P., Lavocat P.,  et al., 2003, A\&A, 411, L141

\bibitem[\protect\citeauthoryear{Lund et al.}{2003}]{lun03}
Lund N., Brandt S., Budtz-Joergesen C., et al., 2003, A\&A, 411, L231


\bibitem[\protect\citeauthoryear{Lyubarskii \& Sunyaev}{1988}]{ls88}
Lyubarskii Yu., Sunyaev R., 1988, Sov. Astron. Lett., 14, 390

\bibitem[\protect\citeauthoryear{Makishima et al.}{1990}]{mak90}
Makishima K., Mihara T., Ishida M., et al., 1990, ApJ, 365, L59

\bibitem[\protect\citeauthoryear{Meszaros \& Nagel}{1985}]{mes85}
Meszaros P., Nagel W., 1985, ApJ, 299, 138

\bibitem[\protect\citeauthoryear{Mihara et al.}{1990}]{mih90}
Mihara T., Makishima K., Ohashi T., et al., 1990, Nature, 346, 250

\bibitem[\protect\citeauthoryear{Mihara et al.}{1998}]{mih98}
Mihara T., Makishima K., Nagase F., 1998, Adv. Space Res., 22, 987

\bibitem[\protect\citeauthoryear{Mowlavi et al.}{2006}]{mow06}
Mowlavi N.,  Kreykenbohm I., Shaw S.E., et al., 2006, astro-ph/0512414

\bibitem[\protect\citeauthoryear{Nakajima et al.}{2006}]{nak06}
Nakajima M., Mihara T., Makishima K., Niko H., 2006, astro-ph/0601491

\bibitem[\protect\citeauthoryear{Okazaki \& Negueruela}{2001}]{oka01}
Okazaki A., Negueruela I., 2001, A\&A, 377, 161

\bibitem[\protect\citeauthoryear{Negueruela et al.}{1999}]{neg99}
Negueruela I., Roche P., Fabregat J., et al., 1999, MNRAS, 307, 695

\bibitem[\protect\citeauthoryear{Pavlov et al.}{1985}]{pav85}
Pavlov G., Shibanov Yu., Silant'ev N., Nagel W., 1985, ApJ, 291, 170 

\bibitem[\protect\citeauthoryear{Pottschmidt et al.}{2005}]{pots05}
Pottschmidt K., Kreykenbohm I., Wilms J., et al., 2005, ApJ, 634, L97 

\bibitem[\protect\citeauthoryear{Revnivtsev et al.}{2004}]{rev04}
Revnivtsev M., Sunyaev R., Varshalovich D., et al., 2004,
Astron. Lett., 30, 382

\bibitem[\protect\citeauthoryear{Stella et al.}{1985}]{st85} Stella L.,
White N.E., Davelaar J., et al., 1985, ApJ, 288, L45

\bibitem[\protect\citeauthoryear{Stella et al.}{1986}]{st86} Stella L.,
White N.E., Rosner R., 1986, ApJ, 308, 669

\bibitem[\protect\citeauthoryear{Swank et al.}{2004}]{sw04} Swank J.,
Remillard R., Smith E., 2004, Astron. Telegram, 349, 1

\bibitem[\protect\citeauthoryear{Terrel \& Predhorsky}{1973}]{tp73}
Terrell J., Priedhorsky W.C., 1973, Bulletin of the American
Astronomical Society, 15, 980

\bibitem[\protect\citeauthoryear{Tsygankov \& Lutovinov}{2005}]{tsy05}
Tsygankov S., Lutovinov A., 2005, Astron. Lett., 31, 88

\bibitem[\protect\citeauthoryear{Winkler et al.}{2003}]{win03}
Winkler C., Courvoisier T.J.-L., Di Cocco G., et al.,  A\&A, 411, L1

\bibitem[\protect\citeauthoryear{White et al.}{1983}]{wh83}
White N., Swank J., Holt S., 1983. ApJ, 270, 711



\end{thebibliography}
\end{document}